\documentclass[traditabstract ]{aa}
\usepackage{natbib,hyperref,wasysym,graphicx,txfonts}
\begin{document}
\def\HEa{HE\,0532-4503}
\def\HEb{HE\,0929-0424}
\def\TONS{TON\,S\,183}
\def\PG{PG\,1232-136}
\def\SB{SB\,410}

\title{A search for radio pulsations from neutron star companions of
  four subdwarf B stars} 
\titlerunning{A search for radio pulsars in four subdwarf B star binaries}

\author{
 Thijs Coenen \inst{1} \and 
 Joeri van Leeuwen  \inst{2,1} \and 
 Ingrid H.\ Stairs \inst{3}
}

\institute{
Astronomical Institute "Anton Pannekoek," University of Amsterdam,
P.O. Box 94249, 1090 GE, Amsterdam, The Netherlands
  \and
Stichting ASTRON, PO Box 2, 7990 AA Dwingeloo, The Netherlands
  \and
Dept. of Physics and Astronomy, University of British Columbia, 6224
Agricultural Road, Vancouver, B.C., V6T 1Z1 Canada
}

\date{}

\abstract{We searched for radio pulsations from the potential neutron
star binary companions to subdwarf B stars \HEa, \HEb, \TONS\ and
\PG. Optical spectroscopy of these subdwarfs has indicated they
orbit a companion in the neutron star mass range. These companions
are thought to play an important role in the poorly understood
formation of subdwarf B stars. Using the Green Bank Telescope we
searched down to mean flux densities as low as 0.2\,mJy, but
no pulsed emission was found. We discuss the implications for each
system.
}


\maketitle

\section{Introduction}\label{section-introduction}
The study of millisecond pulsars (MSPs) enables several types of
  research in astrophysics, ranging from binary evolution
\citep[e.g.][]{2001ApJ...547L..37E}, to the potential detection of
background gravitational radiation using a large set of pulsars
  with stable timing properties \citep{2003ApJ...583..616J}.
Furthermore, high-precision timing provides opportunities for
fundamental physics research: by measuring and modeling pulse arrival
times from binary systems, one can constrain neutron star masses and
equations of state \citep{2010Natur.467.1081D}, or study strong-field
general relativity \citep{1989ApJ...345..434T}.

Finding such millisecond pulsars is done through wide-area surveys or
through targeted surveys, each with specific advantages: with
  limited telescope time, a directed search can afford to dwell on a
  specifically interesting part of the sky for longer than an
  undirected search. A directed search is therefore more sensitive to
  faint pulsars in specific locations than an undirected search would
  be that only incidentally scans over such a location. An undirected
  search however covers a larger area. Directed searches have
targeted sources associated with MSPs, such as globular clusters
\citep{2005Sci...307..892R}, steep spectrum sources
\citep{1982Natur.300..615B}, unidentified Fermi sources
  \citep{2011ApJ...727L..16R} and low mass-white dwarfs
\citep{2007MNRAS.374.1437V,2009ApJ...697..283A}. Here we report the
results of a directed search for radio pulsations from four
short-orbit sub-luminous B dwarf (sdB) binaries identified through
spectroscopy as possibly containing neutron stars
\citep{2008MmSAI..79..608G}.

Sub-luminous B dwarfs are some of the most abundant faint blue objects
\citep{1986ApJS...61..305G}. They are thought to be light (about
$0.5\,M_{\astrosun}$) core helium burning stars with very thin
hydrogen envelopes. After the helium burning in the core stops, sdB
stars evolve directly to the white dwarf cooling sequence \cite[cf. a
  recent review by][]{2009ARA&A..47..211H}.
In their survey of subdwarf B stars \citet{2001MNRAS.326.1391M} find
that $2/3$ of field sdB stars are in short-orbit (P $< \sim$ 10 days)
binaries. Taking into account the insensitivity of their survey to
longer-period binaries, \citet{2001MNRAS.326.1391M} conclude that
binary star evolution is fundamental to the formation of sdB stars. 

Several such binary formation channels have been hypothesized. For an
sdB to form, a light star must lose most of its hydrogen envelope and
ignite helium in its core.
  In these binary systems, sdB stars can be formed through phases of
  Common Envelope evolution where the envelope is ejected, or through
  stable Roche Lobe overflow stripping the donor star of its hydrogen
  envelope. Binary population synthesis models for the above scenarios
  in the case of white-dwarf companions were recently compared by \cite{2003MNRAS.341..669H} and
  \cite{2007A&A...473..569H}.

Specifically interesting for finding new millisecond pulsars, is the
channel that leads to short-period binary systems containing an sdB
star and a neutron star or a black hole \citep{2002ApJ...565.1107P,2010A&A...519A..25G}. Here, the binary
system contains a star that is massive enough to evolve into a neutron
star or a black hole and a secondary, lighter star that is the
progenitor of the sdB star. The secondary is in a wide orbit with a
period of about 20 years. The system undergoes two phases of common
envelope evolution, a supernova and a short X-ray binary phase. A
first common envelope phase starts soon after the primary becomes a
red supergiant and starts overflowing its Roche Lobe. During this
first common envelope phase the orbit of the binary shrinks. After the subsequent
supernova, the primary leaves a neutron star or black hole. 
This neutron star remains visible as a normal radio pulsar for on average $10^{7-8}$ years, and turns off. Once
the secondary evolves off the main sequence the second phase of mass transfer 
starts when the secondary begins to overflow
its Roche Lobe. The system undergoes a short X-ray binary phase, in
which the neutron star is recycled \citep{1991PhR...203....1B}. The second
common-envelope phase starts shortly thereafter, further shrinking the orbit of
the binary and dissipating the envelope of the secondary. The
secondary, which is now mostly stripped of its hydrogen envelope,
ignites helium in its core and continues its evolution as a sdB star
\citep{2002ApJ...565.1107P,2010A&A...519A..25G}. 
From population modeling, \citet{2003ApJ...597.1036P} find that about
1\% of sdB stars evolve as outlined above and orbit a neutron star or
black hole. In such systems the secondary is an sdB star for
$\sim10^8$ years \citep{2009ARA&A..47..211H}, before evolving to a
white dwarf.  

 While {\em low-mass} companions (white dwarfs, main sequence stars) to
  sdB stars have been optically confirmed \citep[e.g.\ ][]{2010A&A...519A..25G},
  the unambiguous identification of a {\em high-mass companion} in the form of a
  radio pulsar would provide further 
constraints on the relative likelihood of different binary evolution
models \citep[as was similarly done for companions of low-mass white dwarfs 
in][]{2007MNRAS.374.1437V}. 
Furthermore, timing of a possible MSP companion
will likely provide an sdB mass measurement, either through the modeling of the orbital
parameters, or more directly and precisely through a Shapiro delay measurement
\cite[e.g.][]{2010ApJ...711..764F}. As millisecond pulsars can turn on shortly 
after
the cessation of accretion \citep{2009Sci...324.1411A} and can shine for more than the
$\sim10^8$ year sdB star lifetime, MSPs will be active throughout the
sdB star phase. Thus, non-detections of radio 
pulsations from these systems could mean the absence of any neutron star, the 
presence of only a very weak radio pulsar, or of a brighter one that is beamed away from Earth. 
Non-detections in a large enough sample of sdB stars eventually provide statistics on
sdB formation channels.

In Section \ref{section-source-selection} we describe how sources
were selected; in Section \ref{section-observations-datareduction}
our observing and data reduction setup is
outlined. Section \ref{section-results} contains our results. In
Section \ref{section-discussion} we discuss these results and compare to
recent optical results on some of our sources.

\section{Source selection}\label{section-source-selection}
 Our source selection was based on the report by
  \cite{2008MmSAI..79..608G} of the detection of the first four sdB
  systems that possibly contain a neutron star or black hole
  companion: \HEa, \HEb, \TONS\ and \PG.

These candidate sdB - neutron star (sdB-NS) systems were identified
through spectroscopic observations of sdB binaries
\citep{2008MmSAI..79..608G}. In sdB-NS systems, the optical spectrum is single 
lined and the compact companion cannot be detected directly optically. 
Assuming tidal
synchronization between the orbital period and the sdB rotational period,
it is possible to constrain the binary inclination from measurements
of the sdB surface gravity, projected rotational velocity and sdB
mass. In their \citeyear{2010A&A...519A..25G} follow-up paper
\citeauthor{2010A&A...519A..25G} give a detailed description of  
this method. In binaries with periods shorter than $1.2$
days the orbit and stellar rotation are found to be
synchronized. Thus, for such systems the companion mass can be derived
from the sdB mass, which is known either from independent mass
measurements \citep[e.g.][]{2008ASPC..392..231F} or inferred from binary
population synthesis \citep{2002MNRAS.336..449H,
  2003MNRAS.341..669H}. Our sources all have orbital periods shorter than these
$1.2$ days. In Table \ref{targets} we describe further target
parameters such as galactic coordinates, local background sky
temperature, distance and estimated dispersion measure. 

\begin{table*}
\begin{center}
\begin{tabular}{l|lllllll}
Object        &  $l$           &   $b$       & $M_{\mathrm{comp}}$ & $P_{\mathrm{orb}}$ &  $T_{sky}$ & $d$    & $DM$ \\
& & & ($M_{\astrosun}$) & (d) & (K) & (kpc) & (pc cm$^{-3}$) \\
\hline
HE 0532-4503  & 251.01     & -32.13 & 1.4 - 3.6 & 0.2656 $\pm$ 0.0001 & 17.0 &  2.8  & 44 \\
HE 0929-0424  & 238.52     &  32.35 & 0.6 - 2.4 & 0.4400 $\pm$ 0.0002 & 15.0 &  1.9  & 38 \\
PG 1232-136   & 296.98     &  48.76 & 2.0 - 7.0 & 0.3630 $\pm$ 0.0003 & 21.9 &  0.57 & 14 \\
TON S 183     & 287.22     & -83.12 & 0.6 - 2.4 & 0.8277 $\pm$ 0.0002 & 19.7 &  0.54 & 12 \\
\end{tabular}
\caption{
Pulsar search targets based on sdB-NS candidate binaries identified by \cite{2008MmSAI..79..608G}.
Companion mass and binary period taken from \cite{2008MmSAI..79..608G} and references therein. 
Sky temperature at 408 MHz extracted from \cite{1982A&AS...47....1H}. Distance determinations taken from 
\cite{2005A&A...430..223L} and \cite{2004A&A...414..181A}. Dispersion measure estimate based
on the NE 2001 electron model \citep{2002astro.ph..7156C}.} 
\label{targets}
\end{center}
\end{table*}

\section{Observations and data reduction}\label{section-observations-datareduction}

On 2008 Oct 15, 18 and 20 we observed these four sdB binaries with the
Robert C. Byrd Green Bank Telescope (GBT).  We used the
lowest-frequency receiver PF1, which provided a bandwidth $\Delta f =
50\,\mathrm{MHz}$ centered around $350\,\mathrm{MHz}$. Every
81.92\,$\mu$s the Spigot backend \citep{kel+05} recorded 2048 spectral
channels in 16-bit total intensity.  
Over the three sessions, three test pulsars (Table \ref{table-test-pulsars}) 
and four sdB-star binaries (Table \ref{table-of-fluxes}) were observed.
Due to scheduling and
weather constraints, integration times of the binaries varied, ranging from 12 to
68 minutes (Table \ref{table-of-fluxes}).

\begin{table}
\begin{center}
\begin{tabular}{l|lllll}
Object &$t_{int}$& $P$ & $DM$ & Peak $S/N$ \\
&(s)& (ms) &(pc cm$^{-3}$)                 \\
\hline
PSR\,J0034-0534 & 60 &   1.877  & 13.8 & 22  \\
PSR\,B0450-18   & 30 & 549.0    & 39.9 & 115 \\
PSR\,B1257+12   & 30 &   6.219  & 10.2 & 15  \\
\end{tabular}
\caption{List of the previously known pulsars that were used as a check of the
telescope, back-end and pulsar-search systems. Test source
PSR\,J0034-0534 is in a short-orbit 1.6-day binary with a $0.16\,M_{\astrosun}$
companion \citep{1994ApJ...425L..41B}. }\label{table-test-pulsars}
\end{center}
\end{table}

  For each system in our sample, we derived the expected dispersion
  measures (DM) from their measured distance, using the NE 2001 free
  electron model \citep{2002astro.ph..7156C}. These expected DMs, listed in Table
  \ref{targets}, are all well below
  $100\,\mathrm{pc\;cm^{-3}}$. Up to $200\,\mathrm{pc\;cm^{-3}}$ the
  trial DMs were spaced such that our search remains sensitive to
  pulsars with rotational periods down to 1 ms; faster pulsars are not
  generally expected \citep{2003Natur.424...42C}. Beyond
  $200\,\mathrm{pc\;cm^{-3}}$ the intra-channel smearing begins to
  dominate, so trial DMs up to $1000\,\mathrm{pc\;cm^{-3}}$ where
created at a lower time resolution.  We next searched these trial DMs
for pulsar signals, in both the time and frequency domain, using the
PRESTO data reduction package \citep{2001PhDT.......123R}.  In the
  time domain we searched for single dispersed pulses of radio emission.
In the frequency-domain search we have to take into account the 
acceleration present in these binary systems. One could potentially derive 
the neutron-star orbital phase from the optical modulation of the sdB star,
and thus estimate the acceleration at the time of the observations. This 
however relies on the assumption that there is no lag between the optical 
light curve and the orbital motion. We have taken the conservative approach 
to search over the full range of possible accelerations throughout the orbit.
As each integration time $t_{int}$ was shorter than
$10\,\%$ of the binary period, our search in period and
period-derivative space sufficed \citep{jk91}.
Using standard PRESTO routines, for each observation the
radio-frequency interference was flagged, candidate period signals were sifted to
remove harmonics, and the top 30 candidates we folded and further
refined by searching nearby dispersion measures, periods and period derivatives 
to maximize signal-to-noise ratio. All candidate plots were
subsequently visually inspected.

At the start of each observing session a known pulsar was observed;
as listed in Table \ref{table-test-pulsars}, these were 
all blindly re-detected by the above sdB-NS search pipeline, thus
  confirming the effectiveness of the search method.

\section{Results}\label{section-results}
No new pulsars were detected toward any of the four sdB stars in our
sample. 

We next investigate how strongly these non-detections rule out the
presence of an MSP. We derive the minimum detectable mean flux density
$S_{min}$ for each observation from the 
pulsar radiometer equation \citep{1985ApJ...294L..25D,1998mfns.conf..103B}:

\[
 S_{min} = \frac{(S/N_{min})T_{sys}}{G\sqrt{n_p t_{int} \Delta
     f}} \sqrt{\frac{W}{P-W}}
 \label{eq-radiometer}
\]

In this equation $S/N_{min}$ is the minimum signal-to-noise ratio at which a 
pulse profile can clearly be recognized as a pulsar profile.
The system temperature $T_{sys}$ is defined as $T_{sys}=T_{sys, GBT}+T_{sky}$ 
where $T_{sys, GBT}=46\,\mathrm{K}$\footnote{\tt{http://science.nrao.edu/gbt/obsprop/GBTpg.pdf}}. For each source,
$T_{sky}$ was extracted from \cite{1982A&AS...47....1H} and scaled
from 408 to 350\,MHz using the $T_{sky} \propto \nu^{-2.6}$ scaling
law from \cite{1987MNRAS.225..307L}. The gain $G$ for the GBT is
$2\,\mathrm{K\,Jy^{-1}}$ \citep{rml06}, the number of polarizations $n_p$, added 
for these total intensity observations, is $2$. Integration time $t_{int}$ and
bandwidth $\Delta f$ for each observation are described in Section 
\ref{section-observations-datareduction}. Finally, $W$
is the pulse width and $P$ is the rotational period. As these are unknown for 
non-detections, we use the average pulse duty cycle $W/P$=$0.12$$\pm$$0.10$ determined
by averaging over the MSP $W_{50}/P$ duty cycles in the ATNF 
pulsar database \citep{2005AJ....129.1993M}. Here and below,
we define MSPs in the ATNF database as those sources with 
periods shorter than $50\,\mathrm{ms}$ (thus, moderately to fully recycled) and magnetic fields lower than 
$10^{11}\,\mathrm{Gauss}$. For each source, we list the resulting $S_{min}$ value in Table \ref{table-of-fluxes}.

  The errors on our values of $S_{min}$ have contributions from the
  error on the average pulse duty cycle and from the systematic error on the minimum signal-to-noise
  ratio $S/N_{min}$. The $S/N_{min}$ at which a pulsar can
  be detected depends on the shape of its profile: a strongly
  peaked profile at high DM is more easily identified as a pulsar than
  a same-S/N sinusoidal profile at low DM, as the latter can also be
  terrestrial interference. We estimate this systematic error 
  as follows. To set a lower limit to the possible $(S/N_{min})$ range we
  inspected the pulse profiles of the test pulsars for increasing
  fractions of the total integration time. We concluded that starting
  from a peak $(S/N_{min})$ of 7 the pulsar re-detections become
  unambiguous. As an upper limit to the range of $(S/N_{min})$ values
  we adopt a value of 10.

To estimate the completeness of our search we derived pseudo
luminosity $L$$=$$Sd^2$ 
limits for each of the binaries in our sample and compared them with the pseudo
luminosities of known MSPs. We define the completeness of our pulsar search
as the percentage of known millisecond pulsars that have pseudo luminosities higher
than our derived pseudo luminosity upper limits, i.e., as the percentage of known 
MSPs that would be detected if placed at the distance of the candidate. We 
searched the literature for distances
$d$ to the
binaries in our sample and used the most recent values. 
For \HEa\ and \HEb\ these are from \cite{2005A&A...430..223L}, 
while the distances to \TONS\ and \PG\ are from \cite{2004A&A...414..181A}, 
where \TONS\ is known as \SB. The former article claims an error on the distance
of $10\%$. In the latter article distance errors are not estimated,
so for those sources we also propagate a 10\% error. All
  distances were derived from models of
sdB atmospheres combined with the magnitude measurement.

\begin{figure}
\includegraphics[width=0.48\textwidth]{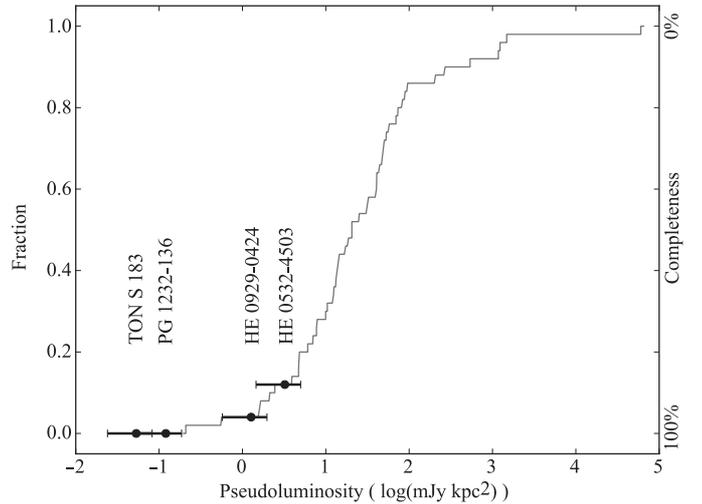}

\caption{Cumulative histogram of pseudo luminosities for the known MSP
  population in the ATNF database. The left axis is labelled with the 
  cumulative fraction of MSPs; the right axis is labelled with survey completeness
  for each candidate.
  The errors on the pseudo luminosity upper limits
  contain errors on the distance, pulse duty cycle and our estimate of
  the relevant range of $(S/N_{min})$ values used with the pulsar
  radiometer equation (see Section \ref{section-results}).}
\label{pseudo-luminosities}
\end{figure}

In Figure \ref{pseudo-luminosities} we compare our limits to the
pseudo luminosities of the millisecond pulsar population. For
  this statistical comparison, we have selected all 50 MSPs with
known $400\,\mathrm{MHz}$ fluxes in the ATNF database
\citep{2005AJ....129.1993M}. We scale these $400\,\mathrm{MHz}$ fluxes
to our central observing frequency of $350\,\mathrm{MHz}$, using the
$-1.6$ average spectral index that \cite{kxl+98}
  find for MSPs. We find that our limits exclude with high confidence
the presence of pulsar emission: in Table \ref{table-of-fluxes} we
list the exact values for this completeness $C$.

\begin{table}
\begin{center}
\begin{tabular}{l|rrrrr}
Object&$t_{int}$&$S_{min}$&$L$&$C$\\
&$\mathrm{(s)}$&$(\mathrm{mJy})$&$(\mathrm{mJy}\;\mathrm{kpc}^{2})$&(\%)\\
\hline
HE 0532-4503&720 & $0.41\pm 0.21$ &   $3.2\pm 1.8   $&88\\
HE 0929-0424&900 & $0.35\pm 0.18$ &   $1.3\pm 0.7   $&96\\
PG 1232-136&1080 & $0.37\pm 0.19$ &  $0.12\pm 0.07  $&100\\
TON S 183&4080   & $0.18\pm 0.09$ &  $0.05\pm 0.03  $&100\\

\end{tabular}
\end{center}
\caption{Integration time $t_{int},$ flux upper limit $S_{min}$ derived from 
    radiometer equation, pseudo luminosity 
    limit $L$ calculated from the flux upper limit, and 
    completeness $C$ compared to the known MSPs in the 
    ATNF database.
}
    \label{table-of-fluxes}
\end{table}

\section{Discussion}\label{section-discussion}
As no radio pulsations were found in any of our observations, we place 
upper limits on the pulsed radio emission from the putative neutron
stars in these systems. Assuming the known MSP luminosity
distribution, we can exclude the presence of such pulsed radio emission
with 100\% certainty for 2 systems: \PG\ and \TONS\ (Figure
\ref{pseudo-luminosities}). 
For 2 other systems, \HEa\ and \HEb, we
were not able to put as strict upper limits on the pseudo luminosity
due to their larger distance and the shorter integration time for
these pointings. For \HEb, our survey would have detected 96\% of MSPs at the \HEb\
distance. For \HEa, this completeness $C$ is 88\% (Table
\ref{table-of-fluxes}, Figure \ref{pseudo-luminosities}).

In relation to these non-detections we now discuss two selection
effects: the fraction of the sdB lifetime during which an MSP is on;
and the beaming fraction, which is the fraction of sky over which an
MSP beam sweeps, and thus from which the MSP is in principle
detectable.  As outlined in Section \ref{section-introduction}, the
MSP and sdB star are formed simultaneously; as MSPs
have ages up of to $10^{9-10}$ years and sdBs have ages up to $10^{8}$
years, an MSP formed in a binary with an sdB star will shine for the
entire age of the sdB, and longer. We thus
conclude there is no age bias against detecting MSPs around sdB
stars. There is however a non-zero chance that for some of our
non-detections a bright MSP emission beam is present, but missing
the Earth. The beaming fraction of MSPs is 0.7$\pm$0.2 \citep{kxl+98};
high compared to normal pulsars, but not unity and thus an important
factor.

So far we have discussed MSPs, as any regular (non-recycled) pulsar
originally present in these systems will have long shut off before the
formation of the sdB began (cf. Section
\ref{section-introduction}). For any non-recycled pulsars that are in
the beam by chance, the $S_{min}$ in Table \ref{table-of-fluxes} list
the minimum detectable flux for that pointing. The pseudo-luminosity
distributions of non-recycled and millisecond pulsars are similar
\citep{2005AJ....129.1993M}, but as the distance to any
chance-coincident non-recycled pulsar is unknown, their pseudo
luminosity is unknown too.

We now compare the results of our search, triggered by the sample from
\cite{2008MmSAI..79..608G}, with the follow-up results of further
spectroscopic investigation recently published in
\cite{2010A&A...519A..25G}. For \PG, where we put a very strict (100\%
complete) limit on the pseudo luminosity, \cite{2010A&A...519A..25G}
now report the mass of the unseen companion $M_{comp}$ to be higher
than $6\;M_{\astrosun}$. This indicates the compact
companion is a black hole; therefore no pulsed radio emission is expected from that
system, in agreement with our non-detection. For \TONS\, the new
spectroscopic results point to a $M_{comp}$ of $0.9\;M_{\astrosun}$
but the error bars allow for a low mass white dwarf (WD). Our
non-detection of pulsed radio emission is compatible with a WD
companion, and since the constraint we put on pulsed radio emission
from \TONS\ is strict (100\% complete) we rule out an MSP beamed
towards Earth.  For \HEb\ a $M_{comp}$ slightly above the
Chandrasekhar limit is reported, but with error bars that also allow
for a high mass WD. Our 96\%-complete non-detection in this case
cannot exclude either possibility but an MSP is unlikely. The
system with the least strict pseudo luminosity upper limit, \HEa, is
reported to contain a companion to the sdB star of $3\;M_{\astrosun}$.
Even if one allows for an sdB star as light as $0.3\;M_{\astrosun}$ the 
companion remains more massive than the Chandrasekhar limit.
Given our
non-detection corresponding to the 12th percentile of the known MSP
population (88\% complete), an MSP is not likely, but this system
remains an interesting candidate for deeper radio
follow-up. Further observations of this source and of new candidate sdB-NS
systems from \cite{2010A&A...519A..25G} are currently ongoing. 

\section{Conclusions}
We have searched for pulsed emission from potential pulsar companions
of four subdwarf B stars. No pulsed emission was found down to
luminosities corresponding to, on average over the 4 sources, the 4th
percentile of the known millisecond pulsar population.

\begin{acknowledgements}
We thank Jason Hessels for help with the search pipeline. This work
was supported by the Netherlands Research School for Astronomy (Grant
NOVA3-NW3-2.3.1) and by the European Commission (Grant
FP7-PEOPLE-2007-4-3-IRG \#224838). Computing resources were provided
by ASTRON. Pulsar research at UBC is supported by an NSERC Discovery
Grant. The National Radio Astronomy Observatory is a facility of the
National Science Foundation operated under cooperative agreement by
Associated Universities, Inc.
\end{acknowledgements}

\end{document}